# Band Structure, Band Offsets, Substitutional Doping, and Schottky Barriers in InSe

*Yuzheng Guo\*[1,2], John Robertson[2]*

[1] College of Engineering, Swansea University, Swansea SA1 8EN, UK
[2] Department of Engineering, Cambridge University, Cambridge, CB2 1PZ, UK
**E-mail:** yuzheng.guo@swansea.ac.uk

**Abstract**. We present a comprehensive study of the electronic structure of the layered semiconductor InSe using density functional theory. We calculate the band structure of the monolayer and bulk material with the band gap corrected using hybrid functionals. The band gap of the monolayer is 2.4 eV. The band edge states are surprising isotropic. The electron affinities and band offsets are then calculated for heterostructures as would be used in tunnel field effect transistors (TFETs). The ionization potential of InSe is quite large, similar to that of $HfSe_2$ or $SnSe_2$, and so InSe is suitable to act as the drain in the TFET. The intrinsic defects are then calculated. For Se-rich layers, the Se adatom is the lowest energy defect, whereas for In-rich layers, the In adatom is most stable for Fermi energies across most of the gap. Both substitutional donors and acceptors are calculated to be shallow, and not reconstructed. Finally, the Schottky barriers of metals are found to be strongly pinned, with the Fermi level pinned by metal induced gap states about 0.5 eV above the valence band edge.

## 1. Introduction

There is presently considerable interest in the layered materials such as the transition metal dichalcogenides (TMDs) and graphene. Graphene is a zero band-gap semimetal with an extremely high carrier mobility. However, the absence of a band gap makes it of minor use for field effect transistors needing power gain. TMDs are true 2D semiconductors with a band gap[1]. However, their carrier mobilities are typically only of order 100 $cm^2$/V.s[2], much less than those of graphene, as their band edge states have d-like character. An advantage of TMDs is that there is a wide range of them with different band gaps and electron affinities[3], so that they could be used in heterostructures in such devices as tunnel field effect transistors.

Black phosphorus is another 2D semiconductor with a band gap, and its carrier mobility of ~1,000 $cm^2$/V.s is much higher than that of TMDs because its band edges have p-like character[4, 5]. However, b-P reacts with water so its devices must be encapsulated. Also, its open lattice allows many substitutional dopants to reconstruct into distorted, non-doping configurations rather than act as shallow substitutional donor or acceptor states[6]. Thus, despite its high carrier mobilities, b-P is not a practical semiconductor.

There is third class of 2D semiconductors, the GaSe and InSe family[3, 7-10]. These consist of a vertically stacked metal-chalcogen double layers with covalent bonding within the layers and van der Waals bonding between the layers. Carrier mobilities in InSe are of order 1000-2000 $cm^2$/V.s[9], similar to those of black phosphorus as they also have s,p-like band edge states. Recently, the quantum Hall effect was observed in the two dimensional electron gas in InSe few layer films[10]. Thus, it is important to understand the electronic structure of InSe, including its band offsets and dopant properties relevant to its use in electronic devices. A device of particular interest is the tunnel field effect transistor (TFET)[11, 12] which is a leading candidate for steep slope transistors. TFETs use a semiconductor heterojunction with a type 2 or 3 band offset. They are typically made from III-V semiconductors, but the lattice matching condition there introduces many interfacial defects which limit performance. TFETs made from stacked layers of 2-dimensional semiconductors do not require lattice matching, and so in principle do not suffer from this type of defect. InSe could be used in such a device with another 2D semiconductor such as WSe$_2$. A final point is that TMD devices tend to be limited by their contact resistances[13, 14]. Thus, we also study the Schottky barrier properties of metals on InSe semiconductors.

## 2. Methods

The calculations are carried out using the plane-wave, density functional code CASTEP[15]. Norm-conserving pseudopotentials are used with a cut-off energy of 500 eV. The In pseudopotential includes the shallow In 4d core levels in the valence shell. Grimme' method[16] is used to correct the density functional theory (DFT) treatment of the van der Waals interactions. The standard DFT underestimates the band gap of semiconductors and insulators. Thus, we use the screened exchange (sX) hybrid density functional[17] to correct these band gap errors. Hybrid functionals such as sX and HSE have been previously applied to the electronic and defect calculations in other 2D systems such as MoS$_2$[18-20].

We use a supercell model for the mono- and few-layer systems, with a vacuum layer of 25 Å between the layer blocks. This is checked to be a good approximation for a 2D slab system. The defect calculations used 120-atom supercells for the few-layer and bulk cases. The minimum distance between mirror image defects is 22 Å. A 2×2×2 Monkhorst-Pack k-point mesh is used for reciprocal space integration in all supercells. The above parameters lead to an energy convergence to less than 0.01eV. The density of states calculations use a k-point mesh of 9x9x1.

For defects, the charge transition states are calculated using the supercell method. Corrections for defect charges and band occupations are applied as in Zunger's scheme[21]. The total energy of the perfect supercell ($E_H$) and the supercell with defect ($E_q$) are calculated for different charge states. The defect formation energy $H_q$ is then found from

$$H_q(E_F) = [E_q - E_H] + q(E_v + \Delta E_F) \quad (1)$$

where $qE_v$ is the change in Fermi energy when charge $q$ is added. No extra correction is needed for a two-dimensional calculation.

## 3. Results and Discussion

3.1 Band structures
Fig 1(a,b) shows the atomic structure of InSe. It consists of four atomic layers in which In occupies 4-fold coordinated sites with In-In bonds and Se occupies 3-fold coordinated sites on the outside of the layers with van der Waals bonding between layers. GaSe and GaS bulk adopts the β-GaSe or 2H structure. On the other hand, bulk InSe tends to adopt the γ or 3R structure. We set all these III-VI compounds in the 2H structure for ease of comparison.

Fig. 2 shows the band structure of an InSe monolayer using the sX functional. The basic ordering of bands is similar to that found by many others[22-32]. The lowest lying Se 4s states are omitted from the figure. The In 4d states lie at -16 eV, and are also off the bottom of the figure. Unlike ML MoS$_2$, the band gap of ML InSe is indirect. The minimum of the conduction band lies at Γ, and the valence band maximum lies just away from Γ in the ΓK direction. This leads to the inverted 'Mexican-hat' valence band structure that has been studied by others[25, 29, 30].

Table 1 gives the calculated band gaps of InSe as a function of the number of layers, in various approximations, first in the generalized gradient approximation (GGA), then SX, and finally in GW from Debbichi et al[27]. As in black phosphorus, the gap $E_g$ depends on the layer number N according to

$$E_g(N) = E_b + (E_1 - E_b)/N \quad (2)$$

where $E_1$ is the gap of the monolayer and $E_b$ is the gap of the bulk.

We notice that the sX band gap is close to the experimental optical/excitonic band gap[10], rather than the quasi-particle gap, as in MoS$_2$[18]. On the other hand, the HSE gap tends to be larger than the optical gap in ML MoS$_2$ [19]. The sX or HSE gap is always less that the quasi-particle gap as given by GW. The difference between the quasi-particle gap and the optical gap is the exciton binding energy, which increases rapidly for small layer numbers due to the poorer screening[33].

3.2 Band character

Each of the bands of GaSe or InSe is known to have a specific atomic and orbital character, as was noted by Schluter[23], Robertson[24], and recently Zolyomi[25]. The main valence band peak consists of Se p$_{x,y}$ states in a bonding combination with In p$_{x,y}$ states. Below this lies a pair of bands at -3 to -6 eV in which the In s states interact with Se p$_z$ states. These states are bonding along both the In-Se bond and the In-In bond. The bands forming the valence band top and the conduction band bottom each consist of In s and Se p$_z$ states. The VB top is anti-bonding along the In-Se bond and bonding along the In-In bond, whereas the CB bottom is anti-bonding along the In-Se bond and along the In-Se bond. Thus, the band gap is a bonding to anti-bonding transition across the In-In bond. This is consistent with the In and Se valences. The Se 4s states are fully occupied. The In and Se p$_{x,y}$ states are half-filled overall. The In 5s states are half-filled and the Se p$_z$ states are filled, by filling only the bonding combination along the In-In bond.

This partial bonding character explains why the band gaps of the GaSe-InSe series vary much less along the series than in the equivalent GaAs-InGaAs compound series. For example, the band gap shrinks from 1.45 eV to 0.33 eV from GaAs to InAs because this is dominated by the lowering of the atomic s state from Ga to In, whereas the gap of GaSe to InSe depends on the much smaller reduction of the gap of the Ga-Ga bond to the In-In bond. The calculated Bader charges for In and Se are +0.15e and -0.15e, which shows that the bonding is not very ionic.

3.3 Band Edge energies and Band Offsets

We have previously calculated the energies of the valence band maximum (VBM) and conduction band minimum (CBM) of the mono-chalcogenides with respect to the vacuum level using a supercell model. The VBM is called the ionization potential (IP) and the CBM is called to electron affinity (EA). A 25 Å vacuum gap is left between the InSe layers. The electrostatic potential is then calculated for each layer normal to the layers, averaged along these layers. The potential within the vacuum gap denotes the vacuum potential. The IP and EA can then be determined with respect to this potential. It is well known that the GGA under-estimates semiconductor band gaps, and that hybrid functionals can be used to correct this error[17, 19]. It is less known that the GGA also makes errors on the IP, and hybrid functionals can also correct these[34]. Sometimes, the exchange mixing in the hybrid functional is varied to fit the band gap[35]. It should be noted that a functional and potential that gives a correct band gap does not necessarily give the correct IP; this must be tested[34]. In particular, it is advisable to use pseudopotentials that retain the shallow d core levels in the valence shell in order to get the correct gap and IP together. This is because these d states interact with the valence band maximum states, and their energy has a strong effect on the VBM energy with respect to the vacuum level.

Fig. 3 shows the VBM and CBM energies calculated using the sX functional, referenced to the vacuum level. They are plotted together with the band energies of some TMDs and of HfS$_2$ [36]. We see that the band edges of the GaSe series lie close to those of HfS$_2$, whereas the MoS$_2$ series lie higher in energy. This follows from HfS$_2$ being a closed shell d0 semiconductor, whereas WSe$_2$ or MoS$_2$ have a d$^2$ configuration with a filled d band, which raises E$_F$ by one band towards the vacuum level.

A tunnel FET requires a type 2 or type 3 band alignment to cutoff the tail of the electron energy distribution and so achieve the small subthreshold slope, below the thermal limit of 60 mV/decade. This can be done using two III-V semiconductors, but this would require a lattice matching of the two semiconductors. The advantage of stacked vdW heterojunctions is that the two semiconductors can be chosen solely in terms of their band gaps and band alignments, while the lattice matching is not necessary because the weak van der Waals bonding does not create defects. Band alignments in the stacked configuration follow the electron affinity rule. It was previously noted that WSe$_2$ with HfS$_2$ or

SnS$_2$ would be a suitable choice for a TFET heterojunction, because the CBM of HfS$_2$ or SnS$_2$ aligns with the VBM of WSe$_2$[3, 36]. However, the acceptor states in HfS$_2$ are not particularly shallow, and HfS$_2$ does not have a high hole mobility, whereas SnS$_2$ has good electronic properties [37] but might not give a single phase under CVD conditions.

We see from Fig 3 that InSe is relatively suitable as a component of a TFET heterojunction. Its band gap is reasonable, its CBM is quite deep below the vacuum level, giving it a type 2 alignment against WSe$_2$. However, its CBM is not quite low enough to show a type 2/3 alignment which would be more desirable. Nevertheless, overall, comparing mobility, synthesis, dopant behavior and band alignment, then InSe is so far the favorable choice for use in TFET heterojunctions.

3.4 Intrinsic Defects

We have also calculated the defect states and formation energies of the In and Se vacancies in InSe. The formation energies are calculated as a function of the Fermi energy $E_F$, following eqn (1). Correction factors must be applied in terms of the band occupations and the charge defects. This is done following the method of Lany and Zunger[21].

Fig 4(a) shows the Se adatom, equivalent to an interstitial, lying on top of another Se. This is the most stable defect for the Se-rich condition. The neutral state has a defect formation energy of 1.23 eV. Fig 4(c) shows an In vacancy. Removing an In from the central In-In dimer allows the Se atoms on both sides to bond to the remaining In atom. The diamagnetic -1 state is the most stable for Fermi energies across most of the band gap, Fig 4(d).

Fig 4(e) shows the Se vacancy. This has an interesting relaxation in which the three surrounding In move closer to each other and forming In-In bonds. For the neutral case, the relaxation is symmetric, but for the -2 case, relaxation is asymmetric, with two short bonds, and one long bond, almost unbonded, because this link is really two filled dangling bond states pointing at each other. Fig 4(g) shows the In 'interstitial' or adatom. This adds on top of the Se-In-In-Se layer, in a hollow site. It is equivalent to the Lewis base noted elsewhere. Overall the Se vacancy is more stable than the In adatom for In-rich conditions. The In adatom is stable in the +1 state for most Fermi energy positions, because this is a diamagnetic state. It is interesting that substitutional donors and acceptors show little reconstruction of their bonding, as noted in the next section, while the native defects show an interesting combination of minor reconstruction plus adatom behavior which is possible because of their layered structure.

3.5 Substitutional Dopants

The n-type dopants are group VII elements at the Se site or group IV elements at the In site. The expected acceptors are group V elements at the Se site, and group II elements at the In site. The doping electrical levels are summarized in Fig 5.

Fig 6(a) shows the relaxed geometry of the neutral substitutional Br donor at the Se site. There is very little lattice relaxation for this defect compared to the undoped lattice. Fig 6(b) shows the formation energy of Br$_{Se}$ as a function of $E_F$. It shows that the +/0 transition lies at 150 mV below the CBM. This is reasonably shallow. It is consistent with the experimentally observed shallow states of donors in bulk InSe.

Fig 6(c) shows the relaxed structure of the substitutional P acceptors at the Se site. It also shows that there is little distortion compared to the undoped lattice. Fig 6(d) shows the defect formation energy vs $E_F$, and the donor transition at an energy of 550 mV above the VBM, a moderately deep acceptor.

Fig 6(e) shows the substitutional Ge donor at the In site. There is no lattice relaxation. The +1/0 transition is calculated to lie at 250 mV below the CBM in Fig 6(f), and is borderline shallow. Fig 6(g) shows the structure of the substitutional Zn at the In site for the monolayer. The -/0 transition is calculated to lie at 550 mV above the VBM.

The binding energies are a function of the dielectric constant and the effective masses. Both the donors and acceptors are known to be shallow in bulk InSe[38-41]. In the monolayers, the dielectric constant declines below the bulk value because of the poorer screening. The effective mass of both holes and electrons is quite low in GaSe and surprisingly isotropic for the layered material[10, 23, 32, 41]. Thus the binding energies are expected to increase above their bulk values, but remain quite shallow. It should be noted that these defect molecule calculations of the binding energies consider only the short range part of the defect potential, and not so much the Coulombic long range part. As such, their accuracy is limited to 0.1 to 0.2 eV.

The key observation for the behavior of the substitutional dopants in InSe is that they remain as dopants, they do not undergo substantial reconstructions like those in b-P, where the dopant sites distort to form non-doping sites obeying the 8-N rule bonding as in amorphous semiconductors[6, 42, 43]. Nor do dopants have the relatively deep character of some dopants in $MoS_2$[44]. These factors are principally because the low effective masses lessen localization effects, which are necessary to give the pseudo-Jahn Teller effects which drives the reconstructions towards the non-doping configurations[45]. Thus, InSe has the advantage that although it is formally a 2D semiconductor, its greater layer thickness and sp bonding gives it the lower effective band mass and shallow dopant characteristics of 3D semiconductors. On the other hand, it retains the advantage of the no lattice-matching condition and low carrier scattering of a 2D semiconductor.

3.6 Contact properties

The device performance of 2D materials is frequently limited by their contact resistances[13, 14]. In some cases this is because the contacts are in the 'on-top' geometry, in others it is because of the presence of Schottky barriers. In order to understand this we have calculated the behavior of ideal interfaces between different metals on InSe in the on-top or 'stacked' geometry following previous work on $MoS_2$ [46-48].

The calculations use supercells with a monolayer of InSe, and five layers of various metals in their FCC phase attached by their (111) faces. The InSe lattice is rotated if necessary to give a lattice matching to the overlying metal. This slab is separated by 25Å of vacuum. The energy difference between the metal Fermi energy and the InSe VBM is then extracted from the partial density of states to define the p-type Schottky barrier height (SBH). There is often considerable hybridization between the InSe states and the metal, so that the InSe VBM is difficult to identify in the combined system[40, 49]. For this purpose, we use the In 4d core level as a reference energy to identify the VBM. The p-type SBH is then plotted versus the experimental work function of the contact metal as in Fig 7[49]. We see that the data follows a linear relation which can be expressed as[50]

$$\phi_n = E_{cnl} - \chi + S (\Phi_M - E_{cnl}) \qquad (3)$$

where $\chi$ is the electron affinity of the semiconductor, $\Phi_M$ is the work function of the metal contract, and S is the Schottky barrier pinning factor, given by $d\phi_n/d\Phi_M$. Within the metal induced gap state model of SBHs, $E_{cnl}$ is the semiconductor's charge neutrality (CNL) reference energy[50], the energy up to which the MIGS are occupied on a neutral surface.

The pinning factor extracted from Fig. 7 is S = 0.18. This relatively low value of S indicates that there is considerable Fermi level pinning by the MIGS. It means that it is not so easy to vary the metal work function to obtain an Ohmic contact.

The charge neutrality level can be calculated from the density of states N(E) as the energy at which the Greens function is zero[50],

$$G(E) = \int \frac{N(E').dE'}{E - E'} = 0 \qquad (4)$$

Here we retain only the In and Se s,p valence states but exclude the In 4d core states, as the latter are too localized to interact strongly with the MIGS. This gives a CNL energy of 0.58 eV for the isotropic case. This is very close to the energy when extracted from the fit of equation (3) to the data points in Fig. 7.

The value S = 0.18 for InSe compares with a value of S = 0.28 calculated for a defect-free $MoS_2$ layer[47]. It is interesting that the theoretical value S = 0.18 is also close to the old experimental value of Kurtin and Mead[51] of S = 0.26 for GaSe. (Note that their slope of barrier height vs Pauling electronegativity, must be converted to the dimension-less barrier height vs electronegativity [49, 52] by the conversion factor of 2.27.) It is also close to the value calculated for the ideal barrier using Monch's[53] dependence of S on optical dielectric constant, S = 0.21. This means that experimental pinning factor for InSe is that for the defect-free interface. This contrasts with the behavior of $MoS_2$ where the experimental pinning factor of S= 0.1[13, 46] is much lower than the theoretical value for the ideal interface [46, 54], and corresponds to pinning by defects[18]. The defects are expected to be chalcogen vacancies, formed by reaction with the contact metal[18, 55]. Given that the formation energy of chalcogen vacancies is similar in each case ( ~ 1.0 eV in the S/Se poor limit) it is unclear why the contact formation process is chemically different in the two cases. On the other hand, there are different and interesting reports on the degree of chemical reaction of metals at GaSe contacts[56, 57] which deserve further attention.

It was noted for $MoS_2$ that the CNL energy depended strongly on the orientation, with the CNL for stacked contacts using the z oriented states being different to that for the lateral contacts which use the x,y states in the in-plane bonds[47]. For this, we calculate the orbital resolved PDOS of InSe monolayers as shown in Fig 8. Applying eqn (2) to the $p_{x,y}$ states and $p_z$,s states separately, we find the stacked $E_{CNL}$ = 0.76 eV for $p_z$,s states, and lateral $E_{CNL}$ = 0.48 eV for the $p_{x,y}$ states. This difference in much less than for $MoS_2$, and originates from the fact that $MoS_2$ has one occupied Mo d–like valence band with particular orientation, whereas InSe has a more of the closed shell isotropic configuration.

## 4. Summary

We have carried out a comprehensive calculation of the electronic structure of monolayer InSe, including its band structure, bonding, optical gaps, band offsets against other 2D layered semiconductors, its native defects, its substitution dopants, and its contact properties. In many ways, monolayer InSe maintains the advantages of three-dimensional (3D) bonded semiconductors with low effective mass band edge states, and shallow dopant sites, while also possessing many advantages of 2D semiconductors such as the ability to form heterostructures of different band gaps, without being subject to the constraints of lattice matching that would hold in normal 3D semiconductors. Thus, it is very suitable for electronic devices such a tunnel FETs.

**Acknowledgements**
This work is supported by the EPSRC grant EP/K016663/1 and EP/P005152/1.

**Table 1.** Calculated band gap values, from GGA and sX methods. Also shown is the GW band gap from Debbichi[27] and the exciton band gap from Bandurin[10].

| Gap (eV) | 1 | 2 | 3 | 4 | 5 | bulk |
|---|---|---|---|---|---|---|
| GGA | 1.44 | 0.97 | 0.81 | 0.73 | 0.68 | 0.49 |
| sX | 2.40 | 1.90 | 1.73 | 1.65 | 1.60 | 1.40 |
| GW | 2.97 | | | | | 1.27 |
| Optical gap (exp) | 2.92 | 1.89 | 1.77 | 1.56 | 1.47 | 1.25 |

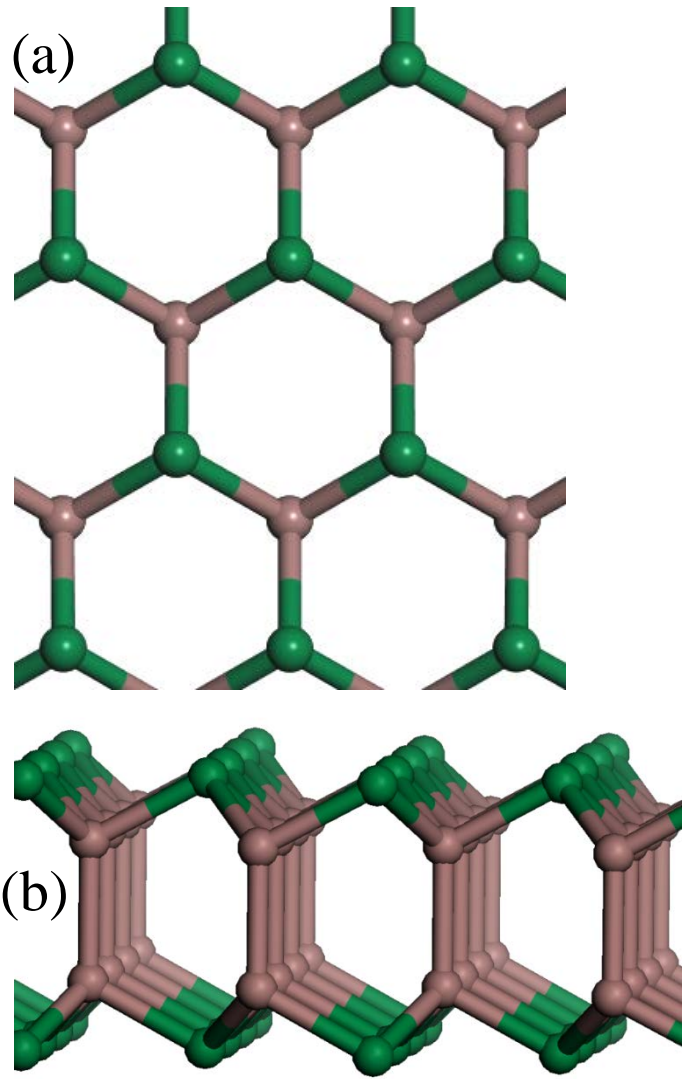

**Figure 1.** Atomic structure of monlayer InSe. In-In bonds form in the central layer to account for the different valence of In and Se. (a) top view (b) side view. In – green. Se – brown.

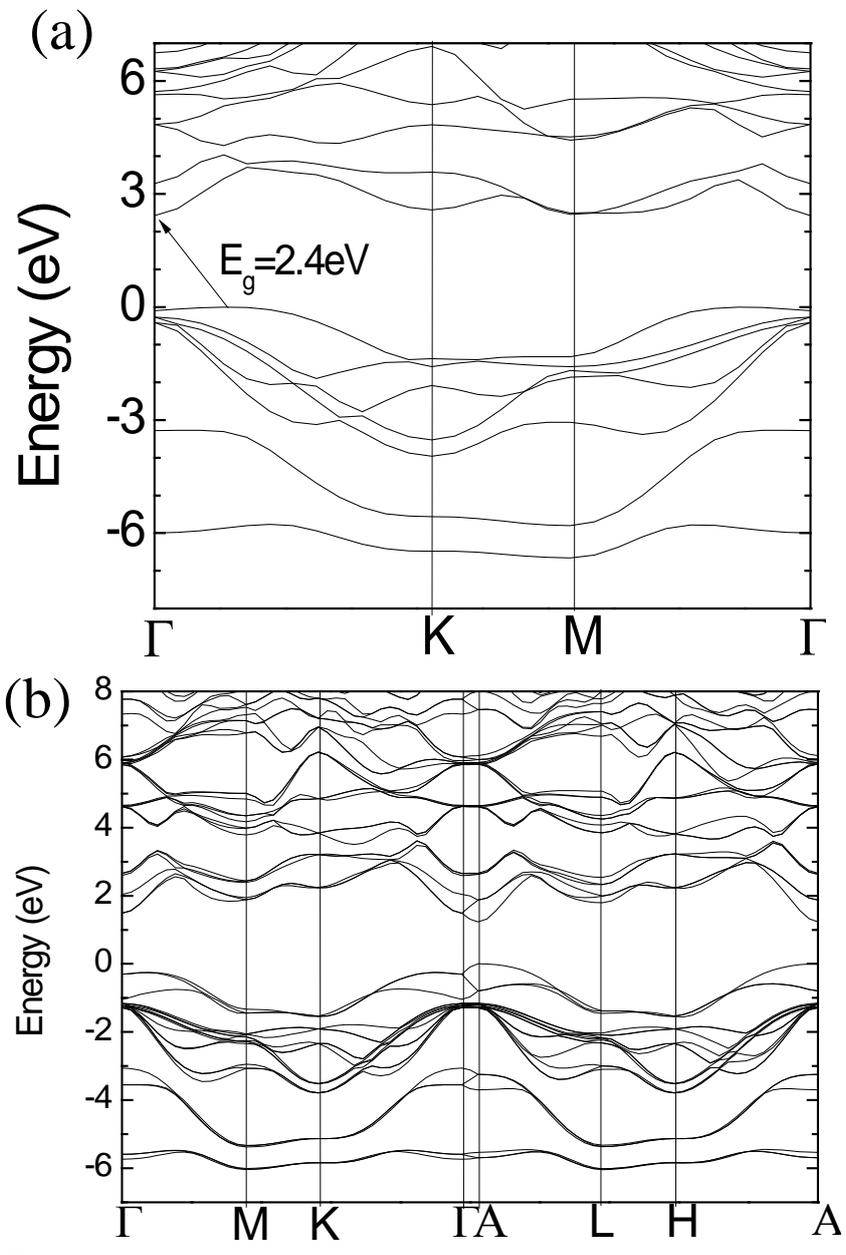

**Figure 2.** Monolayer and bulk band structure MoS$_2$ from sX. The conventional unit cell is used for bulk.

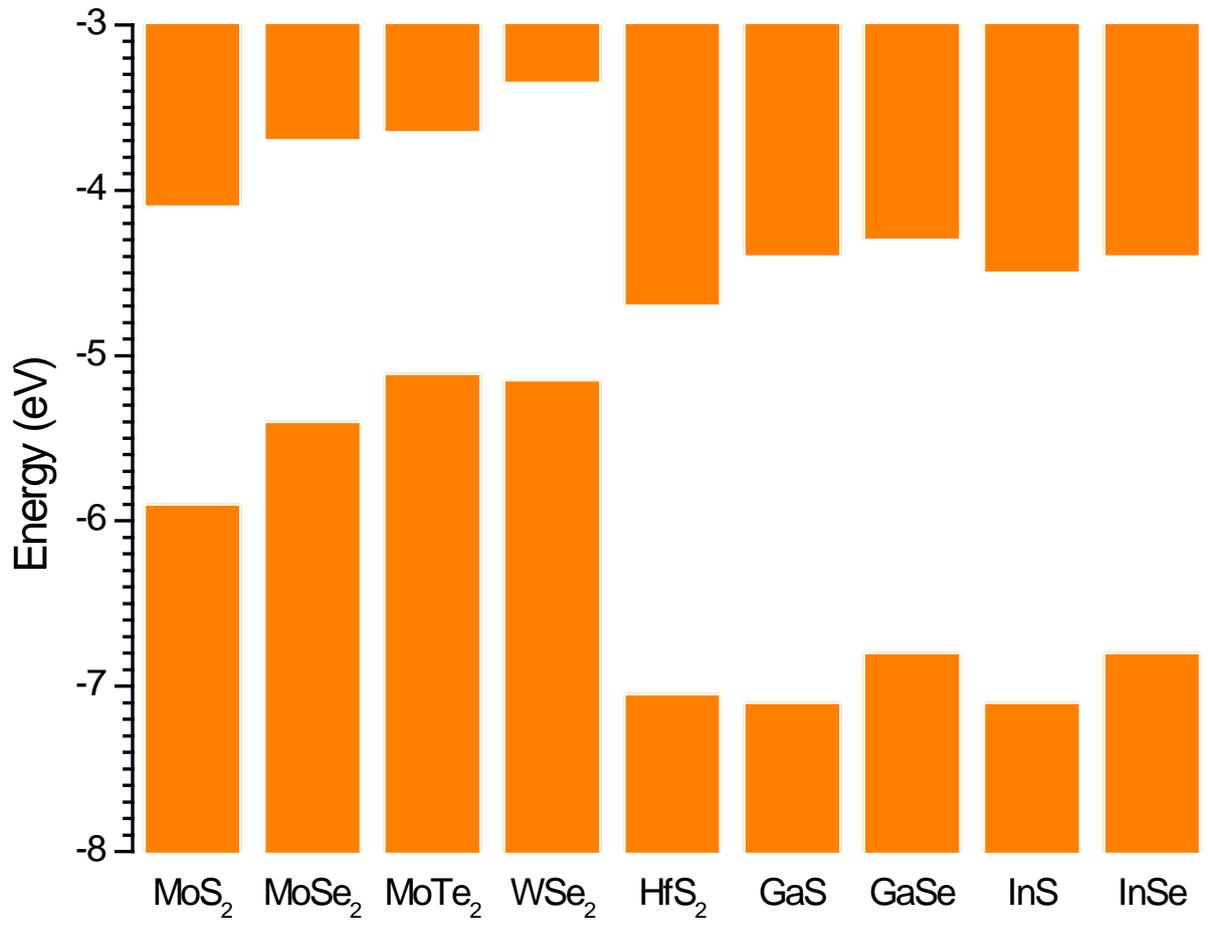

**Figure 3.** Band alignment among various 2D materials. The vacuum level is used to align.

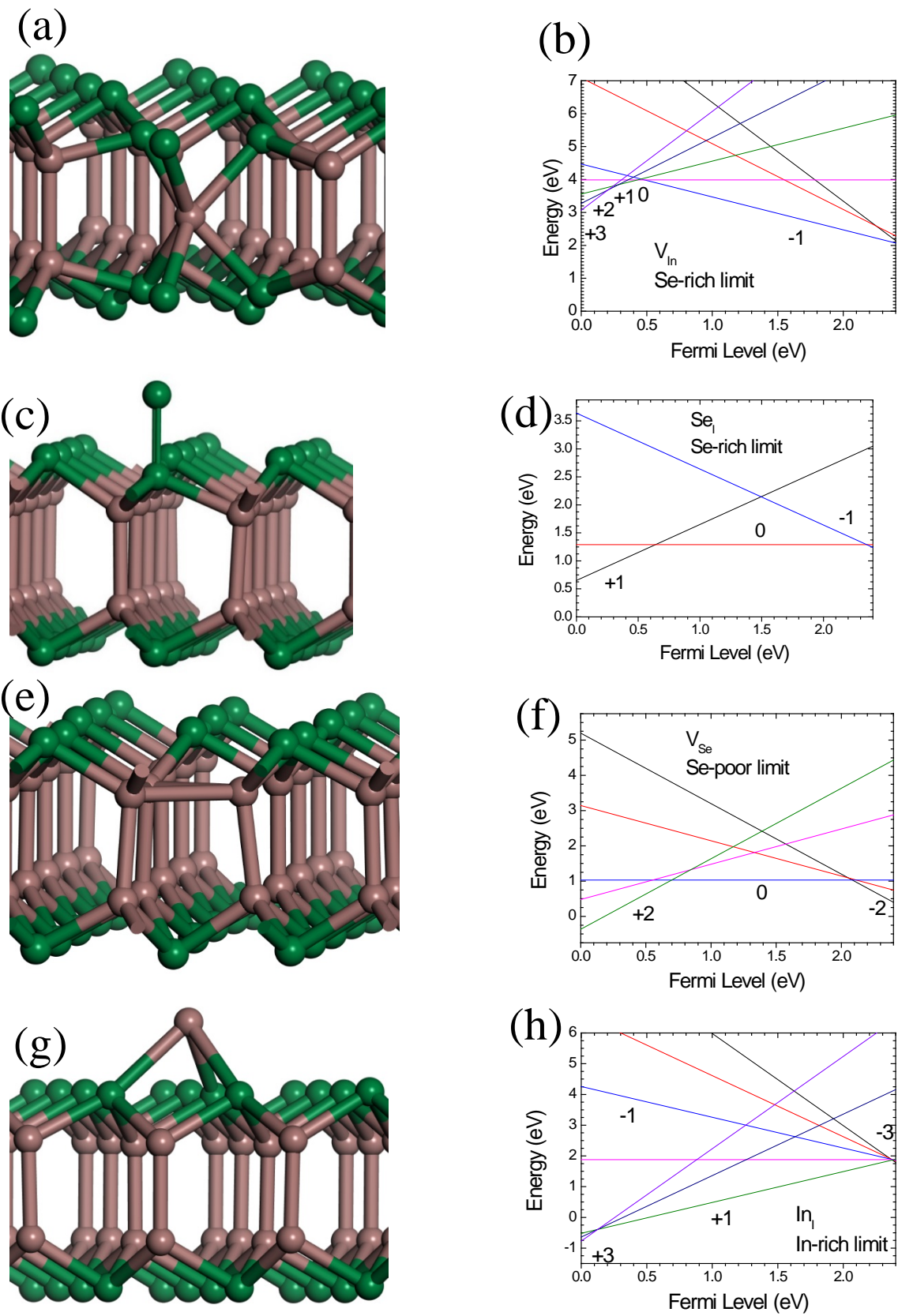

**Figure 4.** Atomic structure of intrinsic defects. (a-b) In vacancy (c-d) Se adatom (e-f) Se vacancy (g-h) In interstitial.

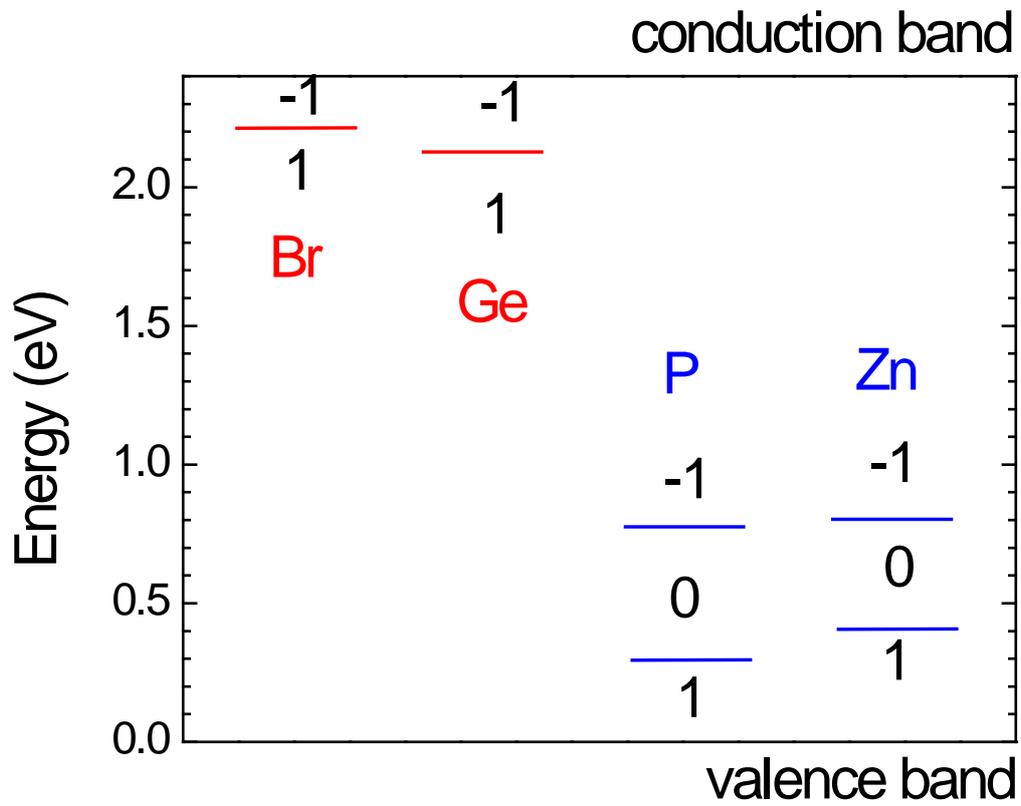

**Figure 5.** The doping defect levels in the band gap. Br and Ge are n-type. P and Zn are p-type.

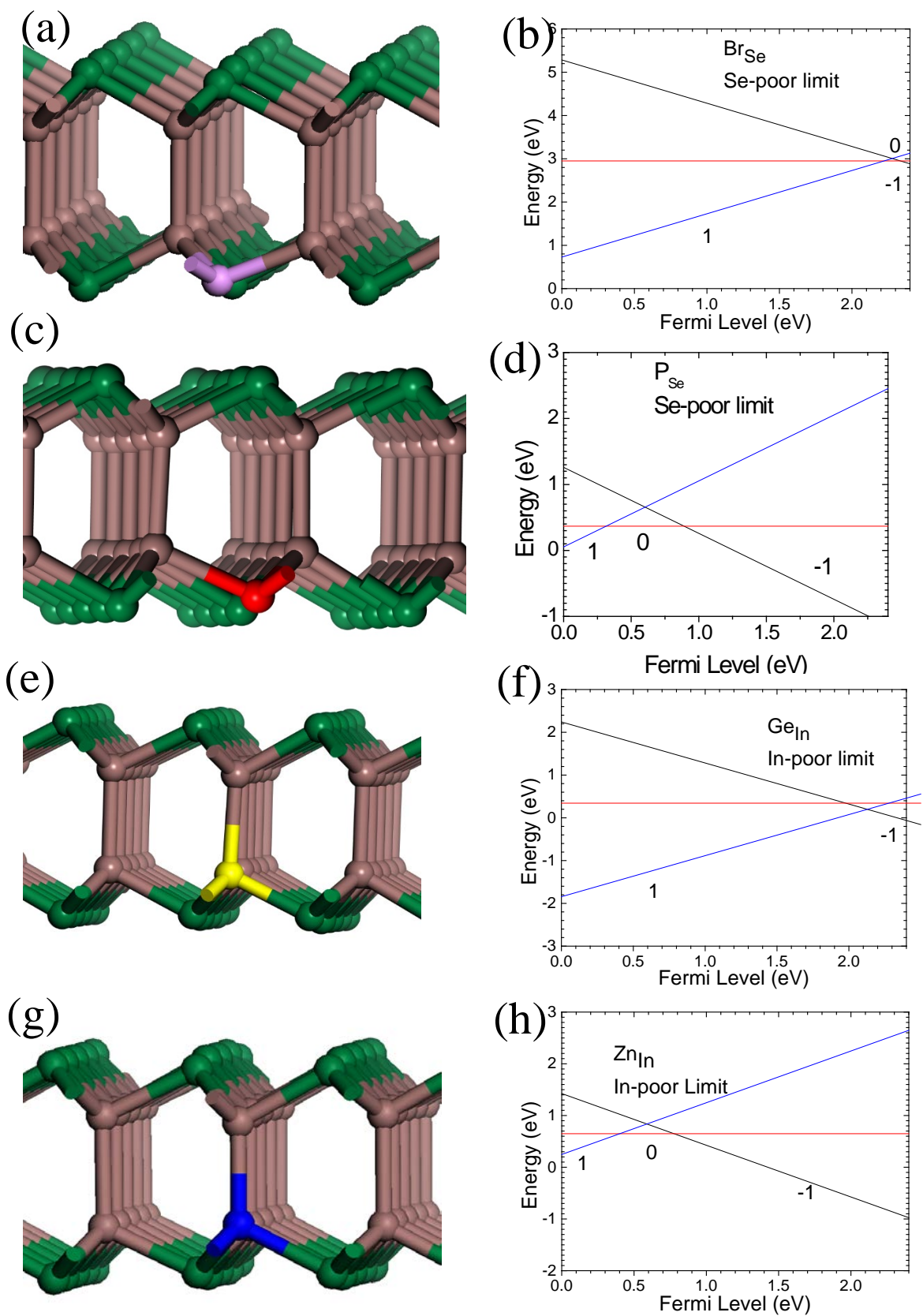

**Figure 6.** Figure caption Fig. 6. Atomic structure and defect transition level of various dopants in InSe. (a-b) Br – pink (c-d) P -red (e-f) Ge – yellow (g-h) Zn – blue.

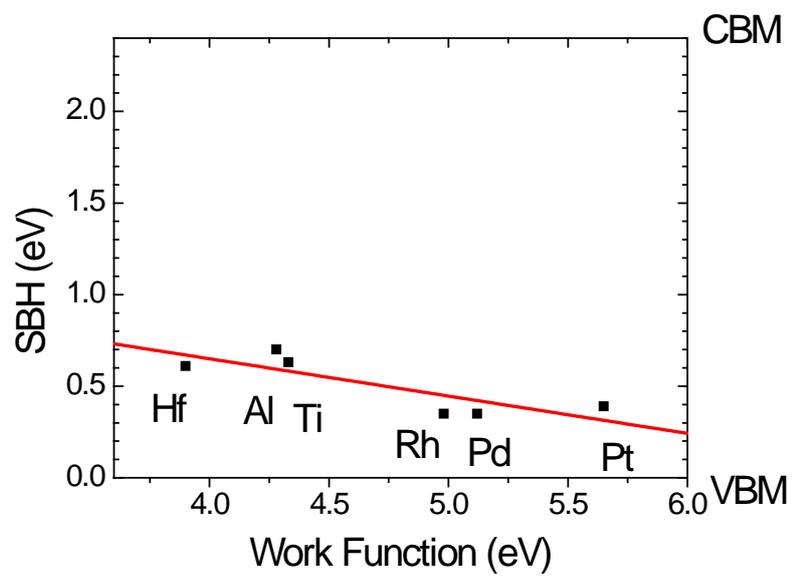

**Figure 7.** Shottky barrier height calculated from supercell model. The VBM is set to be 0eV. The pinning factor is fitted to be is 0.18.

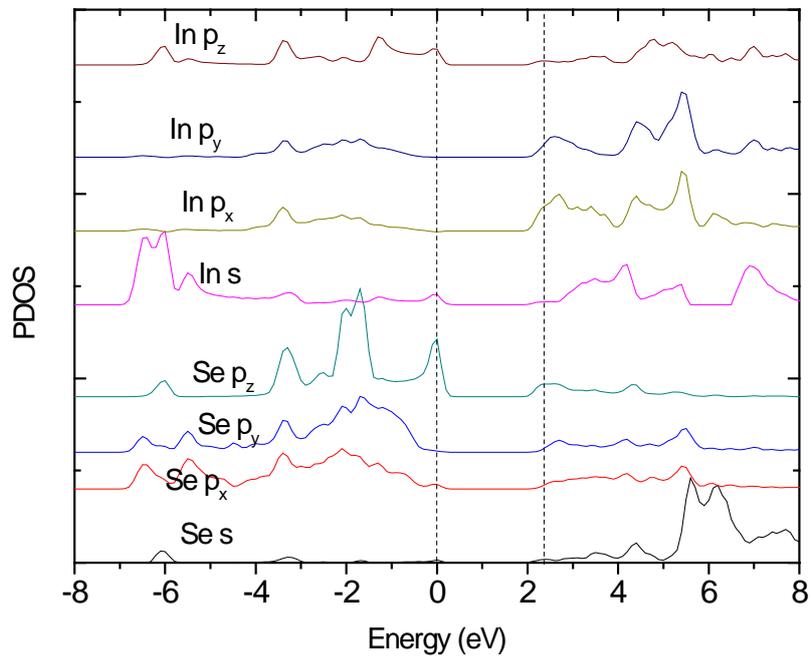

**Figure 8.** Partial DOS of InSe decomposed into angular momentum orbital.

**References**

[1] Mak K F, Lee C, Hone J, Shan J and Heinz T F 2010 *Phys Rev Lett* **105** 136805
[2] Cui X, Lee G H, Kim Y D, Arefe G, Huang P Y, Lee C H, Chenet D A, Zhang X, Wang L, Ye F, Pizzocchero F, Jessen B S, Watanabe K, Taniguchi T, Muller D A, Low T, Kim P and Hone J 2015 *Nat Nanotechnol* **10** 534-40
[3] Gong C, Zhang H J, Wang W H, Colombo L, Wallace R M and Cho K J 2013 *Appl Phys Lett* **103** 053513. 2015 *Appl Phys Lett* **107**, 139904.
[4] Liu H, Neal A T, Zhu Z, Luo Z, Xu X F, Tomanek D and Ye P D 2014 *Acs Nano* **8** 4033-41
[5] Castellanos-Gomez A 2015 *J Phys Chem Lett* **6** 4280-91
[6] Guo Y Z and Robertson J 2015 *Sci Rep-Uk* **5** 14165
[7] Lei S D, Ge L H, Najmaei S, George A, Kappera R, Lou J, Chhowalla M, Yamaguchi H, Gupta G, Vajtai R, Mohite A D and Ajayan P M 2014 *Acs Nano* **8** 1263-72
[8] Feng W, Zheng W, Cao W W and Hu P A 2014 *Adv Mater* **26** 6587-93
[9] Sucharitakul S, Goble N J, Kumar U R, Sankar R, Bogorad Z A, Chou F C, Chen Y T and Gao X P A 2015 *Nano Lett* **15** 3815-9
[10] Bandurin D A, Tyurnina A V, Yu G L, Mishchenko A, Zólyomi V, Morozov S V, Kumar R K, Gorbachev R V, Kudrynskyi Z R, Pezzini S, Kovalyuk Z D, Zeitler U, Novoselov K S, Patanè A, Eaves L, Grigorieva I V, Fal'ko V I, Geim A K and Cao Y 2016 *Nat Nano* **advance online publication**
[11] Ionescu A M and Riel H 2011 *Nature* **479** 329-37
[12] Lu H and Seabaugh A 2014 *Ieee J Electron Devi* **2** 44-9
[13] Das S, Prakash A, Salazar R and Appenzeller J 2014 *Acs Nano* **8** 1681-9
[14] Allain A, Kang J H, Banerjee K and Kis A 2015 *Nat Mater* **14** 1195-205
[15] Clark S J, Segall M D, Pickard C J, Hasnip P J, Probert M J, Refson K and Payne M C 2005 *Z Kristallogr* **220** 567-70
[16] Grimme S 2006 *J Comput Chem* **27** 1787-99
[17] Clark S J and Robertson J 2010 *Phys Rev B* **82** 085208
[18] Liu D, Guo Y, Fang L and Robertson J 2013 *Appl Phys Lett* **103** 183113
[19] Ellis J K, Lucero M J and Scuseria G E 2011 *Appl Phys Lett* **99** 261908
[20] Guo Y, Liu D and Robertson J 2015 *Appl Phys Lett* **106** 173106
[21] Lany S and Zunger A 2008 *Phys Rev B* **78** 235104
[22] Schluter M 1973 *Nuovo Cimento B* **B 13** 313-60
[23] Schluter M, Camassel J, Kohn S, Voitchovsky J P, Shen Y R and Cohen M L 1976 *Phys Rev B* **13** 3534-47
[24] Robertson J 1979 *J Phys C Solid State* **12** 4777-89
[25] Zolyomi V, Drummond N D and Fal'ko V I 2013 *Phys Rev B* **87** 195403
[26] Sun C, Xiang H, Xu B, Xia Y D, Yin J and Liu Z G 2016 *Appl Phys Express* **9** 035203
[27] Debbichi L, Eriksson O and Lebegue S 2015 *J Phys Chem Lett* **6** 3098-103
[28] Ferlat G, Xu H, Timoshevskii V and Blase X 2002 *Phys Rev B* **66** 085210
[29] Rybkovskiy D V, Osadchy A V and Obraztsova E D 2014 *Phys Rev B* **90** 235302
[30] Wickramaratne D, Zahid F and Lake R K 2015 *J Appl Phys* **118** 075101
[31] Brudnyi V N, Sarkisov S Y and Kosobutsky A V 2015 *Semicond Sci Tech* **30** 115019
[32] Zólyomi V, Drummond N D and Fal'ko V I 2014 *Phys Rev B* **89** 205416
[33] Komsa H P and Krasheninnikov A V 2012 *Phys Rev B* **86** 241201
[34] Hinuma Y, Gruneis A, Kresse G and Oba F 2014 *Phys Rev B* **90** 155405
[35] Chen W and Pasquarello A 2012 *Phys Rev B* **86** 035134
[36] Guo Y Z and Robertson J 2016 *Appl Phys Lett* **108** 233104
[37] Huang Y, Sutter E, Sadowski J T, Cotlet M, Monti O L A, Racke D A, Neupane M R, Wickramaratne D, Lake R K, Parkinson B A and Sutter P 2014 *Acs Nano* **8** 10743-55
[38] Mudd G W, Patane A, Kudrynskyi Z R, Fay M W, Makarovsky O, Eaves L, Kovalyuk Z D, Zolyomi V and Falko V 2014 *Appl Phys Lett* **105** 221909
[39] FerrerRoca C, Segura A, Andres M V, Pellicer J and Munoz V 1997 *Phys Rev B* **55** 6981-7

[40] Sanchez-Royo J F, Munoz-Matutano G, Brotons-Gisbert M, Martinez-Pastor J P, Segura A, Cantarero A, Mata R, Canet-Ferrer J, Tobias G, Canadell E, Marques-Hueso J and Gerardot B D 2014 *Nano Res* **7** 1556-68
[41] Ottaviani G, Canali C, Nava F, Schmid P, Mooser E, Minder R and Zschokke I 1974 *Solid State Commun* **14** 933-6
[42] Mott N F 1967 *Adv Phys* **16** 49
[43] Robertson J 1985 *Philos Mag B* **51** 183-92
[44] Dolui K, Rungger I, Das Pemmaraju C and Sanvito S 2013 *Phys Rev B* **88** 075420
[45] Messmer R P and Watkins G D 1973 *Phys Rev B* **7** 2568-90
[46] Guo Y Z, Liu D M and Robertson J 2015 *Acs Appl Mater Inter* **7** 25709-15
[47] Gong C, Colombo L, Wallace R M and Cho K 2014 *Nano Lett* **14** 1714-20
[48] Kang J H, Liu W, Sarkar D, Jena D and Banerjee K 2014 *Phys Rev X* **4** 031005
[49] Michaelson H B 1977 *J Appl Phys* **48** 4729-33
[50] Robertson J 2000 *J Vac Sci Technol B* **18** 1785-91
[51] Kurtin S and Mead C A 1969 *J Phys Chem Solids* **30** 2007-9
[52] Schluter M 1978 *Phys Rev B* **17** 5044-7
[53] Monch W 1987 *Phys Rev Lett* **58** 1260-3
[54] Robertson J, Sharia O and Demkov A A 2007 *Appl Phys Lett* **91** 132912
[55] McDonnell S, Addou R, Buie C, Wallace R M and Hinkle C L 2014 *Acs Nano* **8** 2880-8
[56] Hughes G J, Mckinley A, Williams R H and Mcgovern I T 1982 *J Phys C Solid State* **15** L159-L64
[57] Almeida J, Vobornik I, Berger H, Kiskinova M, Kolmakov A, Marsi M and Margaritondo G 1997 *Phys Rev B* **55** R4899(R)